\documentstyle[12pt]{article}
\begin{document}

\def\RN{Reisner-Nordstr\"om }
\def\SCH{Schwarzchild }
\def\HH{Hartle-Hawking }
\def\BTZ{Ba\~ nados, Teitelboim, Zanelli }
\def\BF{Balbinot-Fabbri }
\def\PoL{Polyakov-Liouville }
\def\PRL{Phys. Rev. Lett. }
\def\be{\begin{equation}} \def\bea{\begin{eqnarray}}
\def\ee{\end{equation}}\def\eea{\end{eqnarray}}
\def\ncr{\nonumber\\ }
\def\sg{\sqrt{-g}}
\def\H{{\cal H}}
\def\ge{g_{cl}}
\def\rzv{{r_*}}
\def\LOG{{\Big( \log{{(r+r_+ )^2(r^2-r_- ^2)\over r^2l^2}}-
{r_- \over r_+ }\log{{r+r_- \over r-r_- }}\Big) }}
\def\CHI{{{3r_+ ^2+r_- ^2\over 3(r_+ ^2-r_- ^2)}\,\log{(r+r_+ )^2\over
r^2-r_- ^2}-{(3r_+ ^2+r_- ^2) r_-\over 3(r_+ ^2-r_- ^2) r_+} \,
\log{r+r_-  \over r-r_- }+{1\over 3}\,\log{(r^2-r_- ^2)^2\over r}
}}

\def\MBVR{Maja Buri\'c
\footnote{E-mail: majab@ff.bg.ac.yu}, Marija Dimitrijevi\'
c\footnote{E-mail: dmarija@ff.bg.ac.yu} and
 Voja Radovanovi\'c\footnote{E-mail: rvoja@ff.bg.ac.yu}\\
{\it Faculty of Physics, P.O. Box 368, 11001 Belgrade,
Yugoslavia}}

% Accents and foreign (in text):

% \let\und=\b                     % bar-under (but see \un below)
% \let\ced=\c                     % cedilla
% \let\du=\d                      % dot-under
% \let\um=\H                      % Hungarian umlaut
% \let\sll=\l                     % slashed (suppressed) l (Polish)
% \let\Sll=\L                     % " L
% \let\slo=\o                     % slashed o (Scandinavian)
% \let\Slo=\O                     % " O
% \let\tie=\t                     % tie-after (semicircle connecting two letters)
% \let\br=\u                      % breve
                % Also: \`        grave
                %       \'        acute
                %       \v        hacek (check)
                %       \~        circumflex (hat)
                %       \%        tilde (squiggle)
                %       \=        macron (bar-over)
                %       \.        dot (over)
                %       \"        umlaut (dieresis)
                %       \aa \AA   A-with-circle (Scandinavian)
                %       \ae \AE   ligature (Latin & Scandinavian)
                %       \oe \OE   " (French)
                %       \ss       es-zet (German sharp s)
                %       \$  \#  \&  \%  \pounds  {\it\&}  \dots

% Abbreviations for Greek letters

\def\a{\alpha}
\def\b{\beta}
\def\c{\chi}
\def\d{\delta}
\def\e{\epsilon}                % Also, \varepsilon
\def\f{\phi}                    %       \varphi
\def\g{\gamma}
\def\h{\eta}
\def\i{\iota}
\def\j{\psi}
\def\k{\kappa}
\def\l{\lambda}
\def\m{\mu}
\def\n{\nu}
\def\o{\omega}
\def\p{\pi}                     % Also, \varpi
\def\q{\theta}                  %       \vartheta
\def\r{\rho}                    %       \varrho
\def\s{\sigma}                  %       \varsigma
\def\t{\tau}
\def\u{\upsilon}
\def\x{\xi}
\def\z{\zeta}
\def\D{\Delta}
\def\F{\Phi}
\def\G{\Gamma}
\def\J{\Psi}
\def\L{\Lambda}
\def\O{\Omega}
\def\P{\Pi}
\def\Q{\Theta}
\def\S{\Sigma}
\def\U{\Upsilon}
\def\X{\Xi}

% Calligraphic letters

\def\ca{{\cal A}}
\def\cb{{\cal B}}
\def\cc{{\cal C}}
\def\cd{{\cal D}}
\def\ce{{\cal E}}
\def\cf{{\cal F}}
\def\cg{{\cal G}}
\def\ch{{\cal H}}
\def\ci{{\cal I}}
\def\cj{{\cal J}}
\def\ck{{\cal K}}
\def\cl{{\cal L}}
\def\cm{{\cal M}}
\def\cn{{\cal N}}
\def\co{{\cal O}}
\def\cp{{\cal P}}
\def\cq{{\cal Q}}
\def\car{{\cal R}}
\def\cs{{\cal S}}
\def\ct{{\cal T}}
\def\cu{{\cal U}}
\def\cv{{\cal V}}
\def\cw{{\cal W}}
\def\cx{{\cal X}}
\def\cy{{\cal Y}}
\def\cz{{\cal Z}}

% Math symbols

\def\bo{{\raise.05ex\hbox{\large$\Box$}\:}}             % D'Alembertian
\def\cbo{{\,\raise-.15ex\Sc [\,}}                       % curly "
\def\pa{\partial}                                       % curly d
\def\de{\nabla}                                         % del
\def\dell{\bigtriangledown}                             % hi ho the dairy-o
\def\su{\sum}                                           % summation
\def\pr{\prod}                                          % product
\def\iff{\leftrightarrow}                               % <-->
\def\conj{{\hbox{\large *}}}                            % complex conjugate
\def\ltap{\raisebox{-.4ex}{\rlap{$\sim$}} \raisebox{.4ex}{$<$}}   % < or %
\def\gtap{\raisebox{-.4ex}{\rlap{$\sim$}} \raisebox{.4ex}{$>$}}   % > or %
\def\TH{{\raise.2ex\hbox{$\displaystyle \bigodot$}\mskip-4.7mu \llap H \;}}
\def\face{\hbox{\normalsize$\;\;\:{\raise.9ex\hbox{\oo n}\mskip-13mu \llap
        {${\buildrel{\hbox{\frtnrm ..}}\over\smile}$}}\:$}}     % happy face
\def\Face{{\raise.2ex\hbox{$\displaystyle \bigodot$}\mskip-2.2mu \llap {$\ddot
        \smile$}}}                                      % another "
\def\dg{\sp\dagger}                                     % hermitian conjugate
\def\ddg{\sp\ddagger}                                   % double dagger
\def\Lhat{{\bf\rlap{\kern-.09em$\hat{\phantom L}$}L}}
\def\Lcheck{{\bf\rlap{\kern-.09em$\check{\phantom L}$}L}}
                        % Also:  \int  \oint              integral, contour
                        %        \hbar                    h bar
                        %        \infty                   infinity
                        %        \sqrt                    square root
                        %        \pm  \mp                 plus or minus
                        %        \cdot  \cdots            centered dot(s)
                        %        \oplus  \otimes          group theory
                        %        \equiv                   equivalence
                        %        \sim                     %
                        %        \approx                  approximately =
                        %        \propto                  funny alpha
                        %        \ne                      not =
                        %        \le \ge                  < or = , > or =
                        %        \{  \}                   braces
                        %        \to  \gets               -> , <-
                        % and spaces:  \,  \:  \;  \quad  \qquad
                        %              \!                 (negative)

% Math stuff with one argument

\def\sp#1{{}^{#1}}                              % superscript (unaligned)
\def\sb#1{{}_{#1}}                              % sub"
\def\oldsl#1{\rlap/#1}                          % poor slash, except for Roman
\def\sl#1{\rlap{\hbox{$\mskip 1 mu /$}}#1}      % good slash for lower case
\def\Sl#1{\rlap{\hbox{$\mskip 3 mu /$}}#1}      % " upper
\def\SL#1{\rlap{\hbox{$\mskip 4.5 mu /$}}#1}    % " fat stuff (e.g., M)
\def\Tilde#1{\widetilde{#1}}                    % big tilde
\def\Hat#1{\widehat{#1}}                        % big hat
\def\Bar#1{\overline{#1}}                       % big bar
\def\bra#1{\Big\langle #1\Big|}                 % < |
\def\ket#1{\Big| #1\Big\rangle}                 % | >
\def\VEV#1{\Big\langle #1\Big\rangle}           % < >
\def\brak#1#2{\Big\langle #1\Big|#2\Big\rangle}         % <|>
\def\abs#1{\Big| #1\Big|}                       % | |
\def\sbra#1{\left\langle #1\right|}             % variable < |
\def\sket#1{\left| #1\right\rangle}             % variable | >
\def\svev#1{\left\langle #1\right\rangle}       % variable < >
\def\sabs#1{\left| #1\right|}                   % variable | |

\def\leftrightarrowfill{$\mathsurround=0pt \mathord\leftarrow \mkern-6mu
        \cleaders\hbox{$\mkern-2mu \mathord- \mkern-2mu$}\hfill
        \mkern-6mu \mathord\rightarrow$}
\def\dvec#1{\vbox{\ialign{##\crcr
        \leftrightarrowfill\crcr\noalign{\kern-1pt\nointerlineskip}
        $\hfil\displaystyle{#1}\hfil$\crcr}}}           % <--> accent
\def\dt#1{{\buildrel {\hbox{\LARGE .}} \over {#1}}}     % dot-over for sp/sb
\def\dtt#1{{\buildrel \bullet \over {#1}}}              % alternate "
\def\ddt#1{{\buildrel {\hbox{\LARGE .\kern-2pt.}} \over {#1}}}% double dot-over
\def\der#1{{\pa \over \pa {#1}}}                % partial derivative
\def\fder#1{{\d \over \d {#1}}}                 % functional derivative
                % Also math accents:    \bar
                %                       \check
                %                       \hat
                %                       \tilde
                %                       \acute
                %                       \grave
                %                       \breve
                %                       \dot    (over)
                %                       \ddot   (umlaut)
                %                       \vec    (vector)

% Math stuff with more than one argument

\def\frac#1#2{{\textstyle{#1\over\vphantom2\smash{\raise.20ex
        \hbox{$\scriptstyle{#2}$}}}}}                   % fraction
\def\ha{\frac12}                                        % 1/2
\def\sfrac#1#2{{\vphantom1\smash{\lower.5ex\hbox{\small$#1$}}\over
        \vphantom1\smash{\raise.4ex\hbox{\small$#2$}}}} % alternate fraction
\def\bfrac#1#2{{\vphantom1\smash{\lower.5ex\hbox{$#1$}}\over
        \vphantom1\smash{\raise.3ex\hbox{$#2$}}}}       % "
\def\afrac#1#2{{\vphantom1\smash{\lower.5ex\hbox{$#1$}}\over#2}}    % "
\def\tder#1#2{{d #1 \over d #2 }}                 % total derivative
\def\partder#1#2{{\partial #1\over\partial #2}}   % partial derivative of
\def\brkt#1#2{{\left\langle #1 | #2 \right\rangle}} % scalar product
\def\secder#1#2#3{{\partial~2 #1\over\partial #2 \partial #3}}  % second "
\def\on#1#2{\mathop{\null#2}\limits~{#1}}       % arbitrary accent (over)
\def\On#1#2{{\buildrel{#1}\over{#2}}}           % alternate "
\def\under#1#2{\mathop{\null#2}\limits_{#1}}    % arbitrary accent-under
\def\bvec#1{\on\leftarrow{#1}}                  % backward vector accent
\def\oover#1{\on\circ{#1}}                              % circle accent

% Young tableaux:  \boxup<a>{\boxes<a>...\boxes<b>}

\def\boxes#1{
        \newcount\num
        \num=1
        \newdimen\downsy
        \downsy=-1.64ex
        \mskip-7.8mu
        \bo
        \loop
        \ifnum\num<#1
        \llap{\raise\num\downsy\hbox{$\bo$}}
        \advance\num by1
        \repeat}
\def\boxup#1#2{\newcount\numup
        \numup=#1
        \advance\numup by-1
        \newdimen\upsy
        \upsy=.82ex
        \mskip7.8mu
        \raise\numup\upsy\hbox{$#2$}}

% Aligned equations

\newskip\humongous \humongous=0pt plus 1000pt minus 1000pt
\def\caja{\mathsurround=0pt}
\def\eqalign#1{\,\vcenter{\openup2\jot \caja
        \ialign{\strut \hfil$\displaystyle{##}$&$
        \displaystyle{{}##}$\hfil\crcr#1\crcr}}\,}
\newif\ifdtup
\def\panorama{\global\dtuptrue \openup2\jot \caja
        \everycr{\noalign{\ifdtup \global\dtupfalse
        \vskip-\lineskiplimit \vskip\normallineskiplimit
        \else \penalty\interdisplaylinepenalty \fi}}}
\def\li#1{\panorama \tabskip=\humongous                         % eqalignno
        \halign to\displaywidth{\hfil$\displaystyle{##}$
        \tabskip=0pt&$\displaystyle{{}##}$\hfil
        \tabskip=\humongous&\llap{$##$}\tabskip=0pt
        \crcr#1\crcr}}
\def\eqalignnotwo#1{\panorama \tabskip=\humongous
        \halign to\displaywidth{\hfil$\displaystyle{##}$
        \tabskip=0pt&$\displaystyle{{}##}$
        \tabskip=0pt&$\displaystyle{{}##}$\hfil
        \tabskip=\humongous&\llap{$##$}\tabskip=0pt
        \crcr#1\crcr}}

% Tables: filler space for tabular*

\def\phil{@{\extracolsep{\fill}}}
\def\unphil{@{\extracolsep{\tabcolsep}}}

% Journal abbreviations

\def\CMP{Commun. Math. Phys.}
\def\NP{Nucl. Phys. B\,}
\def\PL{Phys. Lett. B\,}
\def\PR{Phys. Rev. Lett.}
\def\PRD{Phys. Rev. D\,}
\def\CQG{Class. Quant. Grav.}
\def\IJMP{Int. J. Mod. Phys.}
\def\MPL{Mod. Phys. Lett.}

%\def\ref#1{$\sp{#1]}$}
%\def\Ref#1{$\sp{#1)}$}

% Text style parameters                   textile?

\topmargin=.17in                        % top margin (less 1") (LaTeX)
\headheight=0in                         % height of heading (LaTeX)
\headsep=0in                    % separation of heading from body (LaTeX)
\textheight=9in                         % height of body (LaTeX)
\footheight=3ex                         % height of foot (LaTeX)
\footskip=4ex           % distance between bottoms of body & foot (LaTeX)
\textwidth=6in                          % width of body (LaTeX)
\hsize=6in                              % " (TeX)
\parindent=21pt                         % indentation (TeX)
\parskip=\medskipamount                 % space between paragraphs (TeX)
\lineskip=0pt                           % minimum box separation (TeX)
\abovedisplayskip=1em plus.3em minus.5em        % space above equation (TeX)
\belowdisplayskip=1em plus.3em minus.5em        % " below
\abovedisplayshortskip=.5em plus.2em minus.4em  % " above when no overlap
\belowdisplayshortskip=.5em plus.2em minus.4em  % " below
\def\baselinestretch{1.2}       % magnification for line spacing (LaTeX)
\thicklines                         % thick straight lines for pictures (LaTeX)
\oddsidemargin=.25in \evensidemargin=.25in      % centered margins (LaTeX)
\marginparwidth=.85in                           % marginal note width (LaTeX)

% Paper

\def\title#1#2#3#4{
        {\hbox to\hsize{#4 \hfill  #3}}\par
        \begin{center}\vskip.5in minus.1in {\Large\bf #1}\\[.5in minus.2in]{#2}
        \vskip1.4in minus1.2in {\bf ABSTRACT}\\[.1in]\end{center}
        \begin{quotation}\par}
\def\author#1#2{#1\\[.1in]{\it #2}\\[.1in]}

\def\AMIC{Aleksandar Mikovic\'c
\\[.1in]{\it Blackett Laboratory, Imperial College, Prince Consort Road, London
SW7 2BZ, UK}\\[.1in]}

\def\AMICIF{Aleksandar Mikovi\'c\,
\footnote{Work supported by MNTRS and Royal Society}
\\[.1in] {\it Blackett Laboratory, Imperial College, Prince Consort
Road, London SW7 2BZ, UK}\\[.1in]
and \\[.1 in]
{\it Institute of Physics, P.O. Box 57, 11001 Belgrade,
Yugoslavia} \footnote{Permanent address}\\ {\it E-mail:\,
mikovic@castor.phy.bg.ac.yu}}

\def\AMSISSA{Aleksandar Mikovi\'c\,
\footnote{E-mail address: mikovic@castor.phy.bg.ac.yu}
\\[.1in] {\it SISSA-International School for Advanced Studies\\
Via Beirut 2-4, Trieste 34100, Italy}\\[.1in]
and \\[.1 in]
{\it Institute of Physics, P.O. Box 57, 11001 Belgrade,
Yugoslavia} \footnote{Permanent address}}

\def\AM{Aleksandar Mikovi\'c
\footnote{E-mail address: mikovic@castor.phy.bg.ac.yu}
\\[.1in] {\it Institute of Physics, P.O.Box 57, Belgrade 11001, Yugoslavia}
\\[.1in]}

\def\AMsazda{Aleksandar Mikovi\'c
\footnote{E-mail address: mikovic@castor.phy.bg.ac.yu} and
Branislav Sazdovi\'c \footnote{E-mail:
sazdovic@castor.phy.bg.ac.yu} \footnote{Work supported by MNTRS}
\\[.1in] {\it Institute of Physics, P.O.Box 57, Belgrade 11001, Yugoslavia}
\\[.1in]}

\def\AMVR{Aleksandar Mikovi\'c\,
\footnote{E-mail address: mikovic@castor.phy.bg.ac.yu}
\\[.1in]
{\it Institute of Physics, P.O. Box 57, 11001 Belgrade,
Yugoslavia}
\\[.2in]
Voja Radovanovi\'c \\[.1 in]
{\it Faculty of Physics, P.O. Box 550, 11001 Belgrade,
Yugoslavia}}

\def\AMCVR{Aleksandar Mikovi\'c
\footnote{Permanent address: Institute of Physics, P.O. Box 57,
11001 Belgrade, Yugoslavia}\footnote{E-mail:
mikovic@fy.chalmers.se, mikovic@castor.phy.bg.ac.yu}
\\
{\it Institute of Theoretical Physics, Chalmers University of
Technology,
S-412 96 Goteborg, Sweden}\\[.1in]
and
\\[.1in]
Voja Radovanovi\'c
\footnote{E-mail: rvoja@rudjer.ff.bg.ac.yu} \\
{\it Faculty of Physics, P.O. Box 550, 11001 Belgrade,
Yugoslavia}}

\def\AMVVR{Aleksandar Mikovi\'c
\footnote{On leave from Institute of Physics, P.O. Box 57, 11001
Belgrade, Yugoslavia} \footnote{Supported by Comissi\'on
Interministerial de Ciencia y Tecnologia} \footnote{E-mail:
mikovic@lie1.ific.uv.es}
\\
{\it Departamento de Fisica Te\'orica and IFIC, Centro Mixto
Universidad de Valencia-CSIC, Facultad de Fisica, Burjassot-46100,
Valencia, Spain}
\\[.1in]
Voja Radovanovi\'c
\footnote{E-mail: rvoja@rudjer.ff.bg.ac.yu} \\
{\it Faculty of Physics, P.O. Box 368, 11001 Belgrade,
Yugoslavia}}

\def\AMV{Aleksandar Mikovi\'c
\footnote{On leave from Institute of Physics, P.O. Box 57, 11001
Belgrade, Yugoslavia} \footnote{Supported by Comissi\'on
Interministerial de Ciencia y Tecnologia} \footnote{E-mail:
mikovic@lie1.ific.uv.es}
\\
{\it Departamento de Fisica Te\'orica and IFIC, Centro Mixto
Universidad de Valencia-CSIC, Facultad de Fisica, Burjassot-46100,
Valencia, Spain}}

\def\endtitle{\par\end{quotation}\vskip3.5in minus2.3in\newpage}

% A4

\def\endabstract{\par\end{quotation}
        \renewcommand{\baselinestretch}{1.2}\small\normalsize}

% Letter

\def\xpar{\par}                                         % \par in loops

\def\letterhead{
        \centerline{\large\sf INSTITUTE OF PHYSICS}
        \centerline{\sf P.O.Box 57, 11001 Belgrade, Yugoslavia}
        \rightline{\scriptsize\sf Dr Aleksandar Mikovi\'c}
        \vskip-.07in
        \rightline{\scriptsize\sf Tel: 11 615 575}
        \vskip-.07in
        \rightline{\scriptsize\sf E-mail: MIKOVIC@CASTOR.PHY.BG.AC.YU}}

\def\sig#1{{\leftskip=3.75in\parindent=0in\goodbreak\bigskip{Sincerely yours,}
\nobreak\vskip .7in{#1}\par}}

\def\ssig#1{{\leftskip=3.75in\parindent=0in\goodbreak\bigskip{}
\nobreak\vskip .7in{#1}\par}}

% Referee report

\def\ree#1#2#3{
        \hfuzz=35pt\hsize=5.5in\textwidth=5.5in
        %\begin{document}
        \ttraggedright
        \par
        \noindent Referee report on Manuscript \##1\\
        Title: #2\\
        Authors: #3}

% Book

\def\start#1{\pagestyle{myheadings}\begin{document}\thispagestyle{myheadings}
        \setcounter{page}{#1}}

% Page and section headings and reference stuff

\catcode`@=11

\def\ps@myheadings{\def\@oddhead{\hbox{}\footnotesize\bf\rightmark \hfil
        \thepage}\def\@oddfoot{}\def\@evenhead{\footnotesize\bf
        \thepage\hfil\leftmark\hbox{}}\def\@evenfoot{}
        \def\sectionmark##1{}\def\subsectionmark##1{}
        \topmargin=-.35in\headheight=.17in\headsep=.35in}
\def\ps@acidheadings{\def\@oddhead{\hbox{}\rightmark\hbox{}}
        \def\@oddfoot{\rm\hfil\thepage\hfil}
        \def\@evenhead{\hbox{}\leftmark\hbox{}}\let\@evenfoot\@oddfoot
        \def\sectionmark##1{}\def\subsectionmark##1{}
        \topmargin=-.35in\headheight=.17in\headsep=.35in}

\catcode`@=12

\def\sect#1{\bigskip\medskip\goodbreak\noindent{\large\bf{#1}}\par\nobreak
        \medskip\markright{#1}}
\def\chsc#1#2{\phantom m\vskip.5in\noindent{\LARGE\bf{#1}}\par\vskip.75in
        \noindent{\large\bf{#2}}\par\medskip\markboth{#1}{#2}}
\def\Chsc#1#2#3#4{\phantom m\vskip.5in\noindent\halign{\LARGE\bf##&
        \LARGE\bf##\hfil\cr{#1}&{#2}\cr\noalign{\vskip8pt}&{#3}\cr}\par\vskip
        .75in\noindent{\large\bf{#4}}\par\medskip\markboth{{#1}{#2}{#3}}{#4}}
\def\chap#1{\phantom m\vskip.5in\noindent{\LARGE\bf{#1}}\par\vskip.75in
        \markboth{#1}{#1}}
\def\refs{\bigskip\medskip\goodbreak\noindent{\large\bf{REFERENCES}}\par
        \nobreak\bigskip\markboth{REFERENCES}{REFERENCES}
        \frenchspacing \parskip=0pt \renewcommand{\baselinestretch}{1}\small}
\def\unrefs{\normalsize \nonfrenchspacing \parskip=medskipamount}
\def\Item{\par\hang\textindent}
\def\Itemitem{\par\indent \hangindent2\parindent \textindent}
\def\makelabel#1{\hfil #1}
\def\topic{\par\noindent \hangafter1 \hangindent20pt}
\def\Topic{\par\noindent \hangafter1 \hangindent60pt}

\title{Quantum corrections for BTZ black hole via 2D reduced model }
{\MBVR}{}{July 2001}

\noindent

The one-loop quantum corrections for BTZ black hole are considered
using the dimensionally reduced 2D model. Cases of 3D minimal and
conformal coupling are analyzed. Two cases are considered:
minimally coupled and conformally coupled 3D scalar matter. In the
minimal case, the Hartle-Hawking and Unruh vacuum states are
defined and the corresponding semiclassical corrections of the
geometry are found. The calculations are done for the conformal
case too, in order to make the comparison with the exact results
obtained previously in the special case of spinless black hole.
Beside that we find exact corrections for $\rm{AdS_2}$ black hole
 for 2D minimally coupled scalar field in the Hatrle-Hawking and
Boulware state.

\endtitle

\section{Introduction}

For a long time it was believed that black hole solutions do not
exist in three dimensions, and therefore the discovery of \BTZ
\cite{btz} came as a surprise. This solution has many properties
which the familiar black hole solutions in four dimensions (4D) do
not possess. BTZ black hole can be obtained by identifications of
points in 3D anti-de Sitter (AdS) space \cite{bhtz}, the space of
constant negative curvature. BTZ black hole is locally anti-de
Sitter space, and therefore its singularity is not a curvature
singularity. Obviously, this solution is not asymptotically flat,
although the asymptotic region can be identified. On the other
hand, the fact that BTZ black hole is three dimensional enables
one to work out exactly many computations which  in 4D can be done
only approximately. Among these, the thermal Green function of the
conformally coupled scalar field is found in the framework of the
procedure developed by Avis, Isham and Storey \cite{ais} which
resolves the problem of time-like infinity of AdS. Along with
that, various dimensional reductions from BTZ black hole to two
dimensions are formulated \cite{ao}.

One of the most interesting questions in the analysis of black
holes is the Hawking radiation. A considerable work has been done
in the last couple of years in an effort to find  2D effective
models which can describe the properties of 4D black holes and
radiated field. The main idea of this approach is to consider the
effective action obtained by functional integration of scalar
field as semiclassical correction to the gravitational action.
There are a couple of different variants of 2D effective action
but usually it describes the effects of s-modes of scalar field to
the one-loop order. A similar analysis has been recently extended
\cite{mk} to the reduction of BTZ black hole from three to two
dimensions in the case of minimal 3D coupling with scalar matter.

The purpose of this paper is twofold. Our first goal is to define
the Unruh vacuum by means of dimensionally reduced model. The
definition of the Unruh vacuum  seems still to be an open question
for BTZ black hole. On the other hand, the analogy with \SCH and
\RN case offers a possibility of straightforward generalization of
the nonsingularity of energy-momentum tensor (EMT) on the future
horizon of black hole.

The other important point is to use the advantage of
low-dimensionality of BTZ solution in order to analyze the
dimensional-reduction procedure. This problem is of great
heuristic importance, as dimensional reduction is repeatedly done
in different scenarios of string and brane theories, although the
mechanism is fully understood only at the classical level. The
study of dimensional reduction from four to two dimension in the
case of \SCH  black hole was done previously
\cite{fis,bf,brm,mv4}. There are also some new ideas in the
literature, as dimensional-reduction anomaly \cite{fsz,bffnsz}.
However, the analysis is far from complete. In order to compare
with the results obtained for 3D BTZ \cite{lo,sm,s}, we formulated
a dimensionally reduced theory for the conformal matter. We
defined the \HH vacuum and calculated the backreaction effects.

And finally, for the sake of completeness, we discuss the most
frequently used effective action, \PoL for the case 2D minimally
coupled scalar field. As dimensionally reduced spinless BTZ black
hole is, in fact, two-dimensional black hole with constant
negative curvature, we obtain the full discussion of quantum
corrections of 2D AdS black hole as a subcase.

The plan of the paper is the following. In Sect. 2 we introduce
the general setting of the problem. Sect. 3 gives the analysis of
the Unruh vacuum for the minimally coupled case, while the
conformal coupling is discussed in Sect. 4. Sect. 5 is devoted to
\PoL action and 2D Anti-de Sitter black hole.

\section{General setting}

We start with the three dimensional  gravitational action with
negative cosmological constant ($-2\Lambda =-2l^{-2} <0$) coupled
to the scalar field $f$:

\be \label{S3} \G_ 0^{(3)}={1\over 16\pi G}\int
d^3x\sqrt{-g^{(3)}}\Big(R^{(3)}+{2\over l^2}\Big)-{1\over 16\pi
G}\int d^3x\sqrt{-g^{(3)}}\left( (\nabla f)^2+  \xi R^{(3)} f^2
\right) . \ee The case $\xi =0$ describes the minimal coupling in
3D, while $\xi={1\over 8} $ is the conformal coupling. This action
admits the vacuum solution $f=0$. We consider the BTZ black hole
solution which is locally $\rm{AdS_3}$ space:
 \be \label{g3}
ds_{(3)}^2=-\left( {r^2\over l^2}-lM\right) dt^2+Jldtd\theta
+r^2d\theta ^2 + \left( {r^2\over l^2}-lM+{J^2l^2\over
4r^2}\right)^{-1}dr^2\ . \ee If we construct the metric reduced
from (\ref{g3}) to two-dimensional $t,r$ hypersurface  by the
standard procedure \cite{landau}, we obtain
 \be
ds^2=-\ge dt^2+{1\over \ge}dr^2\ ,\label{g2} \ee where the metric
function $\ge (r)$ is given by \be \ge (r)= {r^2\over l^2}-lM
+{J^2l^2\over 4r^2} \, =\, {(r^2-r_+^2)(r^2-r_-^2)\over r^2l^2} \
.\ee As showed in \cite{bhtz}, quantities $M$ and $J$ have the
meaning of mass and angular momentum. The last equality holds when
$Ml\geq J$; the case $Ml=J$ is the extremal BTZ black hole. One
can see from the Penrose diagram that this space shows great
resemblance with \RN black hole. The outer and inner horizons
$r_\pm$ are given by \be {r_\pm}^2={l^2\over 2}\left( Ml\pm
\sqrt{M^2l^2-J^2}\right)\ . \ee
 Inversely
 \be M={r_+^2+r_-^2\over l^3}\  ,\ \ J={2r_+r_-\over l^2}\ .\ee

 Acchucarro and Ortiz \cite{ao} showed that the metric (\ref{g2})
can be obtained from dimensionally reduced action in the following
way. Let us assume the axially symmetric metric ansatz in three
dimensions: \be \label{ans} ds_{(3)}^2=g_{\mu\nu}dx^\mu dx^\nu
+l^2\Phi ^2(\alpha d\theta +A_\mu dx^\mu )^2 \ ,\ee  where
$g_{\mu\nu}, \Phi , A_\mu $ are two-dimensional metric, dilaton
and $\rm U(1)$ gauge field. All quantities do not depend on
$\theta$. The constant $\a$ will be fixed later.
 3D scalar curvature for anstaz (\ref{ans}) is
\be \label{R3} R^{(3)} =R-{l^2\Phi ^2\over
4}F_{\mu\nu}F^{\mu\nu}-{2\Box \Phi\over \Phi }\ ,\label{reduc}\ee
where $F_{\mu\nu}=\partial _\mu A_\nu -\partial _\nu A_\mu$ while
$R$ is 2D curvature. Also, $\sqrt{-g^{(3)}} = \sg \, l\a \Phi $.
 Introducing the reduction formula (\ref{R3})
into the action (\ref{S3}) and integrating over the angular
variable $\theta$, we obtain 2D action \be \G _0=\G_g +\G_m\ .\ee
Its gravitational part is, up to a total divergence, given by
 \be \label{S2}
\G _g ={l\alpha\over 8G}\int d^2x\sg \Phi \,\left(R-{l^2\Phi
^2\over 4}F^2+{2\over l^2}\right)\ , \ee while the part describing
the matter is \be \label{Sm} \G _m=-{l\alpha\over 8G}\int d^2x\sg
\Phi \left( (\nabla f)^2+\xi f^2 (R-{l^2\Phi ^2\over 4}F^2 -{2\Box
\Phi\over \Phi })\right) \ .\ee In the following, we will choose
$\alpha$ such that ${l\alpha\over 8G}=1$. Also, instead of the
dilaton field $\Phi$, we will use its logarithm $\varphi =\log\Phi
$.

In order to analyze the vacuum fluctuations of the scalar field
$f$, one has to integrate it functionally to the first order in
$\hbar$. Our approximation consists of  the fact that we  do the
functional integration of $f$ in 2D action (\ref{Sm}) and not in
the full 3D action. We use the methods developed in
\cite{mr4,mr3}. The result which we obtained for the one-loop
effective action is \bea  \G_1&=&{1\over 96\pi }\int
d^2x\sg (12\xi-1 )R{1\over \Box} R \nonumber \\
 &+& {1\over 8\pi}\int d^2x\sg \left(
({1\over 4}-2\xi )R{1\over\Box }(\nabla\varphi )^2 + ({1\over
2}-2\xi )R\varphi -{\xi l^2\over
4}R{1\over\Box}e^{2\varphi}F^2\right )  \ .\label{Seff}\eea Note,
that the
 effective actions for 2D dilaton models are analyzed in various
papers \cite{bh,mr4,mr3,no,klv,kv99,klv98,klv99,lmr,no99,nooo99}.

It is easier to use the local form of the action (\ref{Seff}); it
can be obtained by a suitable introduction of
 auxiliary fields \cite{brm,mk}. The local form, however, differs in the cases
 we are going to discuss. Therefore, we proceed with the minimal
 case.

\section{3D minimal coupling}

 For $\xi =0$, the effective action  $\G_1$ can be rewritten in
 the local form as
\be \G _{1,min}  = - {1\over 96\pi}\int d^2x\sqrt{-g}\left( 2R(
\psi -{3\over 2}\chi ) +(\nabla \psi )^2-3(\nabla \psi)(\nabla
\chi )-3(\nabla \varphi)^2\psi -6R\varphi \right) \label{lokS2}
,\ee
 where the auxiliary fields
 \footnote{Note that our auxiliary fields
differ slightly from those introduced in \cite{mk}.}
 $\psi$ and $\chi $ satisfy equations
\be \label{epsi} \Box \psi =R\ , \label{psi} \ee
 \be \Box \chi
=(\nabla \varphi )^2\ . \label{echi} \ee The full semiclassical
action for the minimally coupled field is \bea   \G _{min}& = & \G
_g+\G_{1,min}\ncr &=& \int d^2x\sqrt{-g}e^\varphi \left( R+{2\over
l^2}-{l^2\over 4}e^{2\varphi} F_{\mu\nu}F^{\mu\nu}\right)
\label{celoSm}   \\ & - & \k\int d^2x\sqrt{-g}\left( 2R( \psi
-{3\over 2}\chi ) +(\nabla \psi )^2-3(\nabla \psi)(\nabla \chi
)-3\psi (\nabla \varphi)^2 -6R\varphi \right)\ .\nonumber  \eea We
introduced the constant $\k={1\over 96\pi}$ which will be the
small perturbation parameter in the following. The equations of
motion obtained from the action (\ref{celoSm}), are
(\ref{epsi}-\ref{echi}) and \be \nabla _\mu \left(
e^{3\varphi}F^{\mu\nu}\right) =0 \ ,\label{eF} \ee
 \be
R+{2\over l^2}-{3l^2\over 4}e^{2\varphi}F^2= 6\k
e^{-\varphi}\left( -R+\nabla _\mu (\psi\nabla ^\mu \varphi )
\right) \label{ephi} \ ,\ee
 \bea & &
g_{\alpha\beta} \Box \F -\nabla _\alpha \nabla _\beta \F-\F
g_{\a\b}\Big( {1\over l^2}-{l^2\over 8}\F^2
 F_{\mu\nu}F^{\mu\nu}\Big) - {l^2\over 2}\F^3 F_{\b\m}F^{\
 \mu}
_\alpha \ncr
 & = & T_{\alpha\beta}/2 \ncr
 &=& \k
\Bigl({\nabla _\alpha}\psi{\nabla _\beta}\psi
 -{3\over 2}{\nabla _\alpha}\psi{\nabla _\beta}\chi
-{3\over 2}{\nabla _\alpha}\chi{\nabla _\beta}\psi \ncr
&-&3\psi{\nabla _\alpha }\varphi{\nabla _\beta }\varphi -2{\nabla
_\beta }{\nabla _\alpha} (\psi -3\varphi -{3\over
2}\chi) \label{eg} \\
 &-& {1\over 2}g_{\alpha\beta}\left( (\nabla\psi)^2-3\nabla \psi \nabla
\chi-3\psi (\nabla\varphi)^2\right)+2g_{\alpha\beta}\Box (\psi
-3\varphi -{3\over 2}\chi ) \Bigr). \nonumber\eea

$T_{\alpha\beta} $ is the energy-momentum tensor of the radiated
matter \be T_{\mu\nu}=-{2\over\sg} {\delta\G _1\over \delta
g^{\mu\nu}}\ .\label{EMTdef}\ee

For $\k=0$ we obtain the classical vacuum equations of motion. The
classical solution is \be\Phi =e^\varphi ={r\over l}\,\ ,\ \
F^{\mu\nu}={\epsilon ^{\mu\nu}\over \sqrt{-g}}{Jl\over r^3}=
E^{\mu\nu}{Jl\over r^3}\ ,\ee where $E^{\mu\nu}$ is the covariant
antisymmetric tensor. The zero-th order solution of (\ref{eg}) is
the  BTZ metric (\ref{g2}). Note that $T_{\mu\nu}$ defined in
(\ref{eg}) being of the first order in $\k$, is determined by the
zero-th order solution for the fields $\psi$, $\chi$ and
$\varphi$.

The \HH vacuum state for the minimally coupled scalar field in the
BTZ background in the framework of dimensionally reduced model was
analyzed in details in the work of Medved and Kunstatter,
\cite{mk}. Here we will outline some of their results briefly, in
order to compare them to the other results. In the \HH vacuum all
functions are independent of the time. The solution of
(\ref{epsi}-\ref{echi}) is \be \label{psires}\psi (r)=-\log\ge
(r)+C\rzv \ ,\ee \be \chi (r)=\int {dr\over\ge (r)}\Big(\int dr
{\ge (r)\over r^2}\Big)\,
 +D\rzv \ ,\ee
where the tortoise coordinate $\rzv$  for nonextremal BTZ metric
is given by \be \rzv =\int{dr\over\ge (r)} ={l^2\over
2(r_+^2-r_-^2)}\Big( r_-\log {r+r_- \over r-r_-}-r_+\log {r+r_+
\over r-r_+} \Big)\ .\ee

The assumption that the energy-momentum tensor is regular on the
outer horizon, $r=r_+$, in the freely falling frame  means that
\cite{cf} \be T_{vv}<\infty,\quad {T_{uv}\over\ge }<\infty, \quad
{T_{uu}\over \ge ^2}<\infty \ \ \ {\rm for}\ r=r_+ \ ,\label{reg}
\ee where the components of EMT are given in the null $u,v$
coordinates \footnote {In the further text we will often switch
among the three common choices of coordinates. These are: \SCH
coordinates $t$, $r$, null coordinates $u$, $v$ ($u=t-\rzv$,
$v=t+\rzv$) and Eddington-Finkelstein coordinates $v$, $r$.}.
Using (\ref{reg}), for the constants $C$ and $D$ we obtain
 \be
C=2\,{r_+^2-r_-^2\over l^2r_+},\ \> D=-\, {6r_+^2+2r_-^2\over
3l^2r_+}\ . \label{CD} \ee
 Introducing these values, we  get
 \be \psi (r) =- \log{{(r+r_+ )^2(r^2-r_- ^2)\over r^2l^2}}+
{r_- \over r_+ }\log{{r+r_- \over r-r_- }}\ ,\ee
 \be \chi  (r)=\CHI .\ee
 The corresponding values of EMT in the \HH vacuum are
\bea T_{uu} &=&{\k\over 2 l^4r^6r_+^2} \Bigl( (r-r_+)^2\Bigl(
-3r^6r_+^2-6r^5r_+(2r_+^2-r_-^2) -r^4r_+^2(3r_+^2+2r_-^2)\ncr
 &-&2r^3r_+r_-^2(5r_+^2-3r_-^2)-
3r^2r_+^2r_-^2(2r_+^2-3r_-^2)+10rr_+^3r_-^4+5r_+^4r_-^4 \Bigr)
\ncr &+& 3r_+^2(r^2-r_+^2)^2(r^2-r_-^2)^2\LOG   \Bigr) \ ,
\label{Tuu}
 \eea
\be T_{uv} ={\k\over 2
l^4r^6}(r^2-r_+^2)(r^2-r_-^2)(13r^4+3r^2(r_+^2+r_-^2)-3r_+^2r_-^2)\
, \label{Tuv} \ee \be T_{vv}=T_{uu}\ .\ee

For the energy density of the radiation, $T_{00} =T_{tt}$, we get
 \bea \label{Ttt}
T_{tt}&=&{\k\over l^4r^6r_+}\Bigl( 10 r^8r_+ -6r^7(r_+ ^2-r_-
^2)+8r^6(r_+ ^3-3r_+  r_- ^2)-6r^5(r_+ ^4-r_- ^4)\ncr
 &-&r^4(6r_+ ^5-16r_+ ^3r_- ^2+ 6r_+ r_- ^4)+2r^3r_+ ^2r_-
^2(r_+^2-r_-^2)+2r_+ ^5r_- ^4      \\
 &+&3(r^2-r_+ ^2)^2(r^2-r_-
^2)^2r_+ \LOG \Bigr) \ .\nonumber\eea

There is an important comment on the values of energy density  in
the asymptotic region. One can notice that $T_{tt}$ diverges
asymptotically ($r\to \infty$) as $r^2\log r$, a feature which is
not present in the \SCH case. However, the \SCH metric is
asymptotically flat, while the BTZ metric has nonzero curvature
and $\ge (r)$ behaves like $r^2$ as $r\to\infty$. In order to
understand the properties of the Hawking radiation better, we can
transform to the locally flat coordinates $t^\prime , r^\prime$ at
some distant fixed point $(t,L)$. We get the asymptotics assuming
that $r\sim L\to\infty$. The transformation of coordinates which
we need is \be t^\prime =\sqrt{\ge (L)}t\ ,\ \ r^\prime ={1\over
\sqrt{\ge (L)}}r\ .\label{transf}\ee We see that asymptotically,
$t^\prime \sim{L\over l}t\sim {r\over l}t$, and therefore
 \be T_{{t^\prime t^\prime}}\sim {l^2\over r^2}T_{tt}\sim
 {\k\over l^2}(10+6\log{{r\over l}}) \ ,\ee
 so the energy density diverges logarithmically in the asymptotic
 region. This is a rather unexpected behavior of the minimally
 coupled radiation in BTZ background and we will see that the
 conformal coupling will improve it.

Having fixed the components of EMT, one can find the first
correction of the metric, i.e. solve the equations (\ref{eg}) in
the first order in $\k$. The one-loop corrected static ansatz for
the metric is \be ds^2 = -g(r) e^{2k\omega (r) }dt ^2+{1\over
g(r)}dr ^2 \ .\label{ansg} \ee The function $g(r)$ we take in the
form \be g(r)=\ge (r)\, -\k l m(r)\ ,\ee and the equations
(\ref{eg}) to the first order read: \bea 2{\k\over l}\o^\prime &=&
T_{11}+{T_{00}\over g_{cl}^2}, \label{romega} \\
 \k\,  m^\prime &=&
{T_{00}\over g_{cl}}\ .\ \label{rm}  \eea Their solution  is \bea
m(r)&=&{4r^2-6(r_+ ^2+r_- ^2)\over l^2 r}+{16r_- \over
l^2}\log{{r+r_- \over r-r_- }}                  \\
 &+&{3r^4-r_+ ^2r_- ^2+3r^2(r_+ ^2+r_-
^2)\over l^2 r^3}\LOG\ , \nonumber\eea \bea \omega (r)
&=&F(r)-F(L)\ , \eea where the function $F(r)$ is given by \bea
F(r)&=&l\Big({1\over r}+{(r_+-3r_-)(r_++r_-)\over
r_+(r_+-r_-)(r+r_-)} +{(r_++3r_-)(r_+-r_-)\over
r_+(r_++r_-)(r-r_-)}\ncr &-&{2(3r_+^2+r_-^2)\over
(r+r_+)(r_+^2-r_-^2)}+{32r_+r_-^2\over
(r_+^2-r_-^2)^2}\log(r+r_+)\ncr
 &-&\Big({3r_-\over r_+r}-{8r_-\over (r_++r_-)^2}\Big)\log(r-r_-)+
\Big({3r_-\over r_+r}-{8r_-\over (r_+-r_-)^2}\Big)\log(r+r_-)\ncr
&-&{3\over r}\log{(r+r_+)^2(r^2-r_-)^2\over r^2l^2}\Big)\ .\eea
Here $L$ is the integration  constant. We have assumed that our
system is in a 1D box of size $L$ \cite{fis}.

 The
first correction of the scalar curvature, $R=R_0+\k R_1$, where \be
R_0=-{2\over l^2}-6{r_+ ^2r_- ^2\over l^2r^4}\label{R2}\ ,\ee and
$R_1$ can be expressed in terms of $m$, $\omega$ as
 \be R_1=-3\ge ^\prime \o ^\prime+lm^{\prime \prime}-2\ge \o^{\prime\prime}
 \ ,\ee is regular
on the horizon $r=r_+ $. If one would not have fixed $C$ and $D$
previously, the same values (\ref{CD}) would have been obtained
assuming the regularity of $R_1$  on $ r_+$. We find \bea
R_1&=&{6\over l r_+ r^5}\Bigl( 2r^3(r_+ ^2-r_- ^2) +8r_+ ^3r_-^2
\\ &-&r_+(r^4+r^2(r_+ ^2+r_- ^2)-3r_+ ^2r_- ^2)\LOG \Bigr)\
.\nonumber\eea

The corrected value of the metric gives us the possibility to find
how the horizon of the black hole changes due to the backreaction
of the Hawking radiation. The apparent horizon of the black hole
(which in the static case coincides with the event horizon) in 2D
is defined by \be\label{AHdef}g^{\mu\nu}\partial _\mu r\partial
_\nu r=0\ .\ee In the corrected null coordinates $\bar u$,$\bar
v$, in the general case the metric  is \bea ds^2 & = &
-g(v,r)e^{2\k \omega (v,r)}dv^2+2e^{\k\omega (v,r)}dv\, dr\ncr & =
& -{1\over \mu}g(v,r)e^{2\k\omega(v,r)}d\bar{u}d\bar{v}\ , \eea
with $d\bar v=dv$, $d\bar u=\mu dv-{2\mu\over g(v,r)}e^{-\k\o}dr$
and $\mu$ is the integration factor \cite{mv4}. Analyzing the
condition (\ref{AHdef}) in $\bar u$,$\bar v$ coordinates, we come
to \be\partial _{\bar{u}} r|_{r_{AH}}=0\ ,\ \ \partial _{\bar{v}}
r|_{r_{AH}}=0\ ,\ee which is equivalent to \be
e^{\k\omega}g(v,r)|_{r_{AH}}=0\ . \ee Taking the position of the
apparent horizon in the form \be r_{AH}=r_+ +\k r_1\ ,\ee we get
that the corrected value is \be r_{AH}=r_+
+k\,{l^3m(v,r_+)r_+\over 2(r_+^2-r_-^2)}\ .\label{AH}\ee In the
\HH case (\ref{AH}) gives the one-loop corrected value of the
event horizon: \be r_{AH}=r_+ +\k{lr_+r_-\over
(r_+^2-r_-^2)}\left(  {5r_+^2-r_-^2\over r_+^2} \log {r_+ +
r_-\over r_+ - r_-}+{3r_+^2+r_-^2\over r_+r_-}\log {4(r_+^2 -
r_-^2)\over l^2} - {r_+^2 + 3r_-^2\over r_+r_-}\right)\ . \ee

Having found $\psi$, $\chi$ and the one-loop corrections of the
metric, one can easily calculate the corrected thermodynamical
quantities, temperature and entropy. Entropy is defined as
\cite{w}
$$S=-2\pi \epsilon _{\alpha\beta}\epsilon _{\gamma\delta}{\partial
{\cal L}\over \partial R_{\alpha\beta\gamma\delta}}\Bigg
|_{r_{AH}}\ .$$ For the action (\ref{celoSm}) for entropy we get
\begin{equation}
S=4\pi\left( {r\over l}-\k(2\psi-3\chi-6\log {r\over
l})\right)\Big |_{r_{AH}}\ .
\end{equation}
In \HH state we obtain
\begin{eqnarray}
S & = & 4\pi \Bigg( {r_+\over l}+\k\Big( -{r_+^2+3r_-^2\over
r_+^2-r_-^2}+{5r_+^2-r_-^2\over r_+^2-r_-^2}\log
{4(r_+^2-r_-^2)\over l^2}\\ \nonumber& - & 2{r_+^2+r_-^2\over
r_+^2-r_-^2}\log {r_+^2-r_-^2\over 4r_+^2}+\log
(4r_+(r_+^2-r_-^2))+6\log {r_+\over l}\Big)\Bigg)\ ,
\end{eqnarray}
for entropy while the temperature is given by \be
T_H={r_+^2-r_-^2\over 2\pi l^2r_+}(1-\k F(L)) -\k \Big({r_-^4
+9r_+^4+6r_+^2r_-^2\over 2\pi lr_+^2(r_+^2-r_-^2)}-{8r_-^2\over
 \pi l(r_+^2-r_-^2)}\log{16r_+^2\over l^2}\Big)\ .\ee
  The terms proportional with small parameter $\k$ are one-loop
  corrections for the entropy and Hawking temperature.

We will now analyze the Unruh vacuum. The Unruh vacuum can be
defined as the state of matter whose energy-momentum tensor is
regular on the future event horizon. As it is easily seen, the
region $-\infty <t<\infty$, $r_+\leq r<\infty$ of the $t,r$ plane
transforms into the interior of the triangle $v=-\infty$,
$u=\infty$, $u=v$ in the $u,v$ plane. The line $u=v$ is the
time-like boundary (asymptotic region) of BTZ, $u=\infty$ is the
future event horizon, while $v=-\infty$ is the past event horizon
of BTZ black hole. In order to find the energy-momentum tensor, we
need to solve equations (\ref{epsi}-\ref{echi}) for the general
case. Those equations can be transformed into the system of
partial linear equations which is similar to the one obtained in
\cite{mv4} for the SSG model. For details we refer the reader to
\cite{mv4}. The general solution in the minimal BTZ case reads:
\be \psi (v,r) =-\log\ \ge (r)+\cc (\rzv -{v\over 2}) +\cg (v)\
,\label{gpsi} \ee \be \chi (v,r)=\int {dr\over\ge (r)}\Big(\int
{\ge (r)\over r^2}\, dr\Big) +\cd (\rzv -{v\over 2})+\ch (v) \ ,
\label{gchi}\ee where $\cc ,\cg , \cd , \ch$ are arbitrary
functions of their arguments. Note that the arguments in
(\ref{gpsi}-\ref{gchi}) are written in such a way that the
regularity on the future horizon $u\to \infty$, $v=$const is
equivalent to the regularity on $r\to r_+ $ ($\rzv\to -\infty$),
as the values of $v$ and its functions are constant on the future
horizon.

While the expression for $T_{uv}$ in the general case is the same
as (\ref{Tuv}), for $T_{uu}$ and $T_{vv}$ we obtain \bea
T_{uu}&=&{\k\over 2l^4r^6}\Bigl( -3r^8 +2r^6(r_+ ^2+r_-
^2)-3r^4(r_+ ^4-4r_+ ^2r_- ^2+r_- ^4)-6r^2r_+ ^2r_- ^2(r_+ ^2+r_-
^2)\ncr &+&5r_+ ^4r_- ^4-3(r^2-r_+ ^2)^2(r^2-r_- ^2)^2\bigl( \cg
+\cc -\log{{(r^2-r_+ ^2)(r^2-r_- ^2)\over r^2l^2}}\bigr)\label{TuuU} \\
 &-&\cc
^\prime l^2r^3(3r^4+3r^2(r_+ ^2+r_- ^2)-r_+ ^2r_- ^2)+l^4r^6( {\cc
^\prime}^2 -3{\cc ^\prime}{\cd ^\prime} -2{\cc
^{\prime\prime}}+3{\cd ^{\prime\prime}} )\Bigr)\ ,\nonumber \eea
\bea T_{vv}&=&{\k\over 2l^4r^6}\Bigl( -3r^8 +2r^6(r_+ ^2+r_-
^2)-3r^4(r_+ ^4-4r_+ ^2r_- ^2+r_- ^4)-6r^2r_+ ^2r_- ^2(r_+ ^2+r_-
^2)\ncr &+&5r_+ ^4r_- ^4-3(r^2-r_+ ^2)^2(r^2-r_- ^2)^2\bigl( \cg
+\cc -\log{{(r^2-r_+ ^2)(r^2-r_- ^2)\over r^2l^2}}\bigr)\label{TvvU} \\
&-&2\cg ^\prime l^2r^3(3r^4+3r^2(r_+ ^2+r_- ^2)-r_+ ^2r_-
^2)+4l^4r^6( {\cg ^\prime}^2 -3{\cg ^\prime}{\ch ^\prime} -2{\cg
^{\prime\prime}}+3{\ch ^{\prime\prime}} )\Bigr) \ .\nonumber\eea

From these expressions one can see that in order to inforce the
regularity of $T_{uu}/ \ge ^2$  on the future horizon, one needs
to put the functions $\cc$ and $\cd$ linear in their arguments,
$\cc (x)=Cx$, $\cd (x)=Dx$ with the \HH values of constants $C$,
$D$ given by (\ref{CD}). The functions $\cg$, $\ch$ cannot be
fixed in this manner.  In order to analyze this in more details
let us assume that $\cg$ and $\ch$ are also linear, which is in
accordance with the request of constant luminosity of the black
hole. Under this assumption we see that the difference of outgoing
and ingoing fluxes in the  asymptotic region $r\to\infty$
($\rzv\to 0$ ) has the leading behavior \be T_{uu}-T_{vv}\sim -{\k
r\over 2 l^2}(C-2\cg ^\prime ) \ ,\ee and it is much smaller than
the asymptotic value of the flux \be T_{uu}\sim {3\k r_+ r^2\over
2 l^4 r_+ }\left( 2\log r +{C\over 2}v-\cg -1\right)\ .\ee In
fact, the asymptotic value of the flux is not dominated by the
function $\cg (v)=\cg (t+\rzv )$ for $\rzv\to 0$, although it
fixes the luminosity of the black hole. The dominant term is the
$r^2\log r$-term, and it is the same for $T_{uu}$ and $T_{vv}$.
This is a rather peculiar characteristic of BTZ if we keep in mind
that in the Unruh vacuum for the \SCH black hole the outgoing flux
is asymptotically constant, $T_{uu}\to$ const, while the ingoing
flux vanishes, $T_{vv}\to 0$ as $r\to\infty$ .

One can verify that the given energy-momentum tensor really
describes the Unruh vacuum, because it is regular on the future
horizon but divergent on the past event horizon ($v\to -\infty$,
$u=$ const). If we express EMT (\ref{TuuU}-\ref{TvvU}) in terms of
$r$ and $u$, a logarithmically divergent term for $u=$const,
$\rzv\to -\infty$ appears independently on the choice of the
functions $\cg $ and $ \ch $. I. e., excepting for the case $\cg
(v)={r_+ ^2-r_- ^2\over l^2 r_+}\, v$ which gives the time
independence of EMT and therefore the \HH vacuum state.

Taking the above discussion into account, we conclude that the
functions $\cg$ and $\ch$ cannot be fixed by the properties of EMT
only. The simplest choice for the Unruh vacuum would be $\cg
=\ch=0 $. In that case \be \psi (v,r)=-{r_+ ^2-r_- ^2\over l^2r_+
}\, v-\LOG \ ,\ee
 \bea \chi
(v,r)&=&{3r_+ ^2+r_- ^2\over 3l^2r_+ }\,v+{3r_+ ^2+r_- ^2\over
3(r_+ ^2-r_- ^2)}\,\log{(r+r_+ )^2\over
r^2-r_- ^2}\nonumber \\
&-& {(3r_+ ^2+r_- ^2) r_-\over 3(r_+ ^2-r_- ^2) r_+} \, \log{r+r_-
\over r-r_- }+{1\over 3}\,\log{(r^2-r_- ^2)^2\over r}
 . \eea
 The final expressions for $T_{\mu\nu}$ are
 \bea
T_{uu}&=&{\k\over 2 l^6r^6r_+}\Bigl( l^2 (r-r_+)^2\bigl(
-3r_+r^6-6(2r_+^2-r_-^2)r^5-r_+(3r_+^2+2r_-^2)r^4 \ncr
&-&2r_-^2(5r_+^2-3r_-^2)r^3-3r_+r_-^2(2r_+^2-3r_-^2)r^2+10r_+^2r_-^4r+5r_+^3r_-^4\bigr)
\ncr
 &+&3(r^2-r_+^2)^2(r^2-r_-^2)^2(r_+^2-r_-^2)\, v\label{TUUU}\\
&+&3l^2r_+ (r^2-r_+^2)^2(r^2-r_-^2)^2\LOG \Bigr)\ , \nonumber\eea
\bea T_{vv}&=&{\k\over 2 l^6r^6r_+ }\Big( l^2r_+ \bigl(
-3r^8+2(r_+^2+r_-^2)r^6 \ncr
&-&3(r_+^4+r_-^4-4r_+^2r_-^2)r^4-6r_+^2r_-^2(r_+^2+r_-^2)r^2+5r_+^4r_-^4\bigr)
\ncr &+&3(r^2-r_+^2)^2(r^2-r_-^2)^2(r_+^2-r_-^2)\, v\label{TVVU}\\
 &+&3l^2r_+
(r^2-r_+^2)^2(r^2-r_-^2)^2\LOG \Big)\ .\nonumber\eea The same
values of the energy-momentum tensor are obtained applying the
procedure which is developed by Balbinot and Fabbri, \cite{bf}.

Now we will find the corected geometry. The one-loop ansatz for
the metric is \be ds^2=-g(v,r)e^{2\k
\o(v,r)}dv^2+2e^{\k\o(v,r)}dvdr\ ,\label{anszgu}\ee where
$g(v,r)=g_{cl}(r)-\k lm(v,r)$. Puting this ansatz in the equation
(\ref{eg}) we get \bea {\k\over l}{\partial\omega\over\partial r}
&=& {T_{rr}\over 2}, \label{romegau} \\
 -\k\, {\partial m\over \partial r} &=&
T_{rv}, \label{rmu} \\ \k\, {\partial m\over \partial v} &=&
T_{vv}+\ge (r)T_{vr}\ . \label{vm} \eea

Introducing the values (\ref{Tuv}), (\ref{TUUU}), (\ref{TVVU}) in
the system of equations for $m(v,r)$ and $\omega (v,r)$ we obtain
the one-loop correction for the metric: \bea  m(v,r)&=&
-v\,{r_+^2-r_-^2\over l^4r_+r^3} \left(
-3r^4+8r_+r^3-3(r_+^2+r_-^2)r^2+r_+^2r_-^2\right) \ncr
 &+&{4r^2-6(r_+^2+r_-^2)\over l^2r}+
16{r_-\over l^2}\log{{r+r_-\over r-r_-}}\\
&+&{3r^4+3(r_+^2+r_-^2)r^2-r_+^2r_-^2\over l^2r^3}\LOG \
,\nonumber \label{Unhm} \eea \bea \omega
(v,r)&=&{l(3r_--r_+)(r_++r_-)\over r_+(r+r_-)(r_--r_+)}
-{l(r_--r_+)(r_++3r_-)\over r_+(r-r_-)(r_-+r_+)}
-{2l(3r_+^2+r_-^2)\over (r+r_+)(r_+^2-r_-^2)} \nonumber \\
&+&{8lr_-\over (r_+^2-r_-^2)^2}\left( (r_+^2+r_-^2)\log{r-r_-\over
r+r_-}+2r_+r_-\log{(r+r_+)^2\over r^2-r_-^2}\right)\\ &-&{3l\over
r}\LOG \ncr &+&{l\over r}-3v{r_+^2-r_-^2\over lr_+r}\ .\nonumber
 \label{Unhomega} \eea

The value for the apparent horizon in this case is
 \bea r_{AH} & =
& r_+ +\k {l\over r_+(r_+^2 - r_-^2)}\Bigl( r_+(3r_+^2+r_-^2)\log
{4(r_+^2 - r_-^2)\over l^2} \\ & - & r_-(r_-^2-5r_+^2)\log {r_+ +
r_-\over r_+ - r_-}-r_+(r_+^2+3r_-^2)-{v\over
l^2}(r_+^2-r_-^2)^2\Bigr)\nonumber \ .\eea Entropy for Unruh state
is given by
\begin{eqnarray}
S & = & 4\pi \Bigg( {r_+\over l}+k\Big(-{r_+^2+3r_-^2\over
r_+^2-r_-^2}+{4r_+v\over l^2}+{5r_+^2-r_-^2\over r_+^2-r_-^2}\log
{4(r_+^2-r_-^2)\over l^2}\nonumber\\ & + & 2{r_+^2+r_-^2\over
r_+^2-r_-^2}\log {4r_+^2\over r_+^2-r_-^2}+\log
(4r_+(r_+^2-r_-^2))+6\log {r_+\over l}\Big)\Bigg)\ .
\end{eqnarray}

\section{Conformal coupling}

We will now discuss the case of the conformally coupled matter.
The coupling constant for the conformal coupling in three
dimensions is $\xi ={1\over 8}$. The local form of the effective
action (\ref{Seff}) for this value is \be \G_{1,conf}={\k\over
2}\int d^2x\sqrt{-g}\left( R( 2\psi +\chi ) +(\nabla \psi
)^2+(\nabla \psi)(\nabla \chi )-{3l^2\over 4}\psi e^{2\varphi}F^2+
6R\varphi \right) \ ,\ee and the full action reads \bea \G
_{conf}&=& \G _g+\G_{1,conf}\ncr & = & \int d^2x\sqrt{-g}e^\varphi
\left( R+{2\over l^2}-{l^2\over 4}e^{2\varphi}
F_{\mu\nu}F^{\mu\nu}\right)     \label{Sconf} \\
 &
+ & {\kappa\over 2} \int d^2x\sqrt{-g}\left( R( 2\psi +\chi )
+(\nabla \psi )^2+(\nabla \psi)(\nabla \chi )-{3l^2\over 4}\psi
e^{2\varphi}F^2+ 6R\varphi \right) \ .  \nonumber \eea The
equations which follow from the variational principle for
(\ref{Sconf}) are \be \label{Epsi} \Box\psi =R\ ,\ee
\be\label{Echi} \Box\chi =-{3l^2\over 4}e^{2\varphi}F^2 \ ,\ee
\be\label{EF} \nabla _\mu \Big( (1+\frac{3}{2}\kappa\psi
e^{-\varphi})e^{3\varphi}F^{\mu\nu}\Big) =0\ ,\ee \be\label{Ephi}
R+{2\over l^2}-{3l^2\over 4}e^{2\varphi}F^2=-\kappa
e^{-\varphi}\left( 3R-{3l^2\over 4}\psi e^{2\varphi}F^2 \right)\ee
and
 \bea & & g_{\alpha\beta} \Box \F -\nabla _\alpha \nabla _\beta \F-\F
g_{\a\b}\Big( {1\over l^2}-{l^2\over 8}\F^2
 F_{\mu\nu}F^{\mu\nu}\Big) - {l^2\over 2}\F^3 F_{\mu\beta}F^\mu
_{\ \alpha} \ncr
 & = & T_{\alpha\beta}/2 \ncr
 &=& -{\kappa\over 2} \Bigl({\nabla _\alpha}\psi{\nabla _\beta}\psi
 +{1\over 2}{\nabla _\alpha}\psi{\nabla _\beta}\chi
+{1\over 2}{\nabla _\alpha}\chi{\nabla _\beta}\psi\nonumber\\
&-&{3l^2\over 2}\psi e^{2\varphi}F_{\beta\n}F _\alpha^{\ \n}
-{\nabla _\beta }{\nabla _\alpha }(2\psi +\chi +6\varphi )
 \label{Eg} \ncr
 &-& {1\over 2}g_{\alpha\beta}( (\nabla\psi)^2+\nabla \psi \nabla
\chi  -{3l^2\over 4}\psi e^{2\varphi}F^2) +g_{\alpha\beta}\Box
(2\psi +\chi +6\varphi  ) \Bigr)\ .  \eea

We can again take that the solution of (\ref{Ephi}) for dilaton is
$e^\varphi ={r\over l}$, and this in fact represents our choice of
the radial coordinate. Then (\ref{EF}) can also be solved exactly
\be F^{\mu\nu}=E^{\mu\nu}e^{-3\varphi}{J\over
l^2}(1+3\kappa{l\psi\over 2r})^{-1}\ \label{F}.\ee

We proceed with the static case in order to find the values of
fields in thermal equilibrium. The zero-th order solution for
$\psi$ is, as before \be \psi (r)=-\log\ge (r) +C\rzv\ ,\ee while
for $\chi$ we have \be \chi (r)=\int {dr\over \ge (r)}\Big( \int
{3J^2l^2\over 2r^4}dr\Big) +D\rzv\ . \ee

Our goal is to solve the equation (\ref{Eg}) determining the
backreaction to the metric, i.e. to extract the equations for the
functions $m(r)$ and $\omega (r)$ from it. Let us note that, as it
can be seen from  (\ref{F}), in the conformal  case the
"electromagnetic field" $F_{\mu\nu}$  has the corrections of the
first order in $\kappa$. This means that in three dimensions the
angular part of the metric has also to be corrected. Technically,
there are the first-order terms on the both sides of equation
(\ref{Eg}). We will collect all first-order terms on the right
hand side. Then the equations for the metric read  \bea 2{\k\over
l}\o^\prime &=& T_{11}+{T_{00}\over g_{cl}^2}, \label{romegac} \\
 \k\,  m^\prime &=&
{T_{00}\over g_{cl}}-{3\k \over 2}{J^2l^2\psi\over r^4}\ ,\
\label{rmc} \eea
 under the same ansatz (\ref{ansg}) for $g_{\mu\nu}$ as before.

 The procedure to determine the integration
constants is as for the minimal coupling. The values of constants
for the \HH vacuum are \be C=2\, {r_+^2-r_-^2\over l^2r_+}\ ,\ \
D={2r_- ^2\over l^2r_+}\ \ .\ee

For the auxiliary field $\chi$ we get \bea\chi (r)&=&{r_+^2\over
r_+^2-r_-^2}\log{{(r+r_-)(r- r_- )\over r^2}} \ncr &+& {r_-^3\over
r_+(r_+^2-r_-^2)}\log{{(r+r_-)\over (r-r_-)}}- {2r_-^2\over
r_+^2-r_-^2}\log{{r+r_+\over r}} \ ,\eea while $\psi$ is the same
as in 3D minimal case  (and as it will be for the \PoL action).
The energy-momentum tensor in the \HH vacuum reads:
 \bea T_{uu}=T_{vv}&=&-\k{(r-r_+ )^2\over 2 l^4r^6}\Bigl(
3r^6+6r^5r_++r^4(3r_+^2-10r_- ^2) \nonumber \\
 &-&20r^3 r_+r_-^2+3r^2r_- ^2(-3
r_+ ^2+r_- ^2)+8r r_+r_-^4+4 r_+^2r_- ^4 \Bigr) \ ,
\label{Tuuc}\eea \bea T_{uv}&=&\k{(r^2- r_+^2)(r^2-r_- ^2)\over 2
l^4r^6}\Bigl( r^4+3r^2( r_+^2+r_- ^2)-12r_+ ^2r_- ^2\ncr &-&3 r_+
^2r_- ^2\LOG\Bigr) \ ,\label{Tuvc}\eea while the energy density is
\bea T_{tt}&=&-{\k\over l^4r^6}\Bigl(2r^8-4r^6(2 r_+ ^2+3r_-
^2)+r^4(6 r_+^4+38 r_+^2r_- ^2+6r_- ^4)\ncr
 &-&2r^3 r_+ r_-^2( r_+^2 -r_-^2 )-24r^2 r_+ ^2r_- ^2( r_+ ^2+r_- ^2)+16 r_+ ^4r_-
 ^4\Bigr)\\
&-&3\k {(r^2- r_+ ^2)(r^2-r_- ^2)\over  l^4r^6} r_+ ^2r_- ^2\LOG \
. \nonumber \label{Tttc}\eea We see now that the asymptotic
behavior of EMT is improved, as the leading term for $r\to\infty$
is $T_{tt}\sim -{\k r^2\over l^4}$. This means that the energy
density of radiation in the locally Minkowskian frame is constant.
The solution for the functions $m$ and $\omega$ also turns out to
be nonsingular on the horizon $r=r_+$. It is given by: \bea
m(r)&=&{1\over l^2r^3}\Bigl( -2r^4-6r^2( r_+ ^2+r_- ^2)+6 r_+
^2r_- ^2-2r_-r^3\log{{r+r_- \over r-r_- }} \nonumber \\
 &-& r_+ ^2r_- ^2\LOG \Bigr)\ ,
 \label{mc}\eea
\be \o(r)=F(r)-F(L)\ , \label{omegac}\ee where $F(r)$ is given by
\bea F(r) & = & 4{l\over r}-{l(r_+-2r_-)\over
2(r+r_-)(r_+-r_-)}-{l(r_++2r_-)\over
2(r-r_-)(r_++r_-)}+{lr_-^2\over (r+r_+)(r_+^2-r_-^2)} \nonumber \\
& - & {2lr_+r_-^2\over (r_+^2-r_-^2)^2}\log {(r+r_+)^2\over
r^2-r_-^2}+{lr_-(r_+^2+r_-^2)\over (r_+^2-r_-^2)^2}\log
{r+r_-\over r-r_-} .\eea For the first correction of the curvature
we obtain \be R_1= {6\over lr^5}\left( r^4+3 r_+ ^2r_- ^2-2
r_+^2r_- ^2\LOG \right)\ .\label{Rc}\ee

We can now compare our results with results in the literature. The
Green functions for BTZ black hole were calculated in
\cite{lo,sm,s}. The starting point of this calculation is the
Green function for the scalar field in $\rm{AdS_3}$ space.
However, as AdS space has a time-like infinity, it does not have a
Cauchy surface. The prescription to fix the boundary conditions
for the wave equation and define the orthonormal basis of
eigenfunctions for the quantization is the following \cite{ais}.
One conformally maps AdS into the half of the Einstein static
universe (ESU), which is spatially compact and has a well defined
Cauchy problem. The solutions for the conformally coupled scalar
field in ESU can be mapped back into the solutions for the
conformally coupled scalar field in AdS, and hence from the basis
of eigenfunctions in ESU one inherits the basis in AdS. The use of
the complete basis in ESU gives the so-called "transparent
boundary conditions". Transparent boundary conditions have the
feature that the energy of  scalar field is not conserved. Also,
it is possible to define two types of "reflective boundary
conditions" (Dirichlet and Neumann), such that the energy in both
cases is conserved. The final step of the construction of Green
functions for BTZ black hole is to apply the method of images.

The Green function for spinning BTZ black hole for the transparent
boundary conditions is given by Steif \cite{s}, the backreaction
to the metric was discussed by Martinez and Zanelli \cite{mz}. We
will not compare our results to those, as the transparent boundary
conditions are not appropriate for description of the \HH state
because of nonconservation of energy. Lifschytz, Ortiz \cite{lo},
and Shiraishi, Maki \cite{sm} found the Green functions for
reflective boundary conditions in the spinless case, $J=0$. In
both of these papers some aspects of the behavior of the
energy-momentum tensor and of backreaction effects were extracted
and we will quote shortly keeping in mind that our results,
obtained by dimensional reduction, are approximate.

The expectation value of the energy-momentum tensor for the
spinless BTZ black hole is given in \cite{lo,sm}. Even in the
spinless  case the components of EMT have relatively complicated
form of infinite sum and nonpolynomial behavior, so it is not easy
to compare them with results
 (\ref{Tuuc}-\ref{Tuvc})
which look much simpler. In \cite{lo} was shown that the energy
density is positive for Dirichlet boundary conditions, while for
Neumann boundary conditions it is not. We obtained $T_{tt}\sim
-{\k r^2\over  l^4}$ for $r\to \infty$, or in the locally flat
frame, $T_{{t^\prime}{t^\prime}}\sim -{\k \over l^2}$. However, we
know from the analysis of the \SCH case that the dimensional
reduction can change the sign of the energy, as it takes into
account not all but only a part of the modes of scalar field.  EMT
is regular for $r=r_+$ and singular as $r\to 0$ both in
\cite{lo,sm} and in our calculation.

Since the metric ansatz in \cite{lo} is not the same as the one we
used and it does not seem to be correct \cite{mz}, we will compare
the corrections for the curvature , only. In \cite{lo} was
obtained that the curvature scalar $R^2$ diverges like ${1\over
r^6}$ near $r=0$. For $J=0$ in (\ref{Rc}) we see that $R_1={6\over
lr}$. However, this is only the correction of the two-dimensional
 piece of the curvature scalar. In order to find the full
 three-dimensional
correction, we should employ the reduction formula (\ref{reduc}).
Using the solutions written to the first order in $\kappa$ \be\Phi
={r\over l}\ ,\ \ F^2=-2{J^2l^2\over r^6}(1-3\kappa{\psi l\over
r})\ ,\ee and \be \Box\Phi ={1\over\sg}\partial _\mu (\sg
g^{\mu\nu}\partial _\nu\Phi )={1\over l}(\ge ^\prime  +\kappa\ge
\omega ^\prime -\kappa lm^\prime )\ ,\ee we find the first
correction of 3D curvature: \be R_1^{(3)}={8\over lr}+6{Ml^2\over
r^3}\ .\ee From this expression it can be seen that $R^2$ also
diverges like $1/r^6$ near $r=0$. Note that in the zero-th order,
the reduction formula gives $R_0^{(3)}=-{6\over l^2}$, while
$R_0=-{2\over l^2}-{3J^2l^2\over 2r^4}$. We will later use the
fact that for $J=0$ we get $\rm{AdS_2}$ black hole.

Finally, we can compare the metric corrections which are in
\cite{sm} given in  the large mass limit. The function $\mu (r)$
used in \cite{sm} is proportional to our $m(r)$. It behaves as
 \be
\mu (r)\sim {r_+\over r}-1\ ,\ \ {\rm \ Neumann\ b.c.}\ee \be \mu
(r)\sim\Big({r_+\over r}\Big)^3-2{r_+\over r}-1\ ,\ \ {\rm \
Dirichlet\  b.c.}\ ,\ee while our result for $m(r) $ is \be
m(r)=-2{r^2+3{r_+}^2\over l^2r}\ .\ee If the limit $M\to\infty$
can be understood as $r_+\gg r$,
 than the behavior of $m(r)$ is the same as the one obtained in
 \cite{sm}
for the Neumann boundary conditions (up to an integration
constant which we, for the sake of simplicity, discarded in the
expression (\ref{mc}) for $m$).

Coming back to the 2D conformal matter model, we want to add some
 remarks, skipping the details of calculations. It is always
interesting to give a particular analysis of the extremal black
hole, and this was done in \cite{mk} for the case of minimally
coupled matter. The conclusion was that BTZ black hole behaves
similarly to dimensionally reduced \RN black hole \cite{mv3}.
Namely, the \HH  EMT for extremal black hole is different from the
limit $r_+\to r_-$ of nonextremal black hole. E. g., it behaves
differently on the event horizon: while we have the regularity
 for the extremal black hole, it is not present
in the nonextremal limit. Surprisingly, this is not so in the
conformally coupled case: nonextremal and extremal black holes
behave similarly. One can check that exactly $r_+\to r_-$ ($C\to
0$) gives the best regularity properties to the energy momentum
tensor.

The other peculiar thing for the conformal case is that one cannot
define the Unruh vacuum  obeying all regularity conditions on the
future horizon, as it was possible for the minimal case.

\section{2D minimal coupling}

In this section we will consider minimal coupled scalar field to
gravity in two dimensions. Performing the functional integration
of 2D scalar field in the path integral we will obtain \PoL
effective action. It is very often used for the exact or
qualitative description of one-loop quantum effects of the scalar
field. This action  was widely discussed in the context of string
theory and 2D dilaton gravity and it is given by \be \G
_{1,PL}=-{1\over 96\pi}
 \int d^2x\sqrt{-g} R{1\over\Box}R\ ,\ee
or in the local form \be \G _{1,PL}=-\k
 \int d^2x\sqrt{-g}\left( (\nabla \psi)^2+2R\psi \right)\ .\label{SPL}\ee
 An auxiliary scalar field $\psi$ satisfies the equation $\Box\psi =R$.
  The energy-momentum
 tensor determined by (\ref{SPL}) is
 \be T_{\mu\nu}=2\k \left({\nabla _\mu}\psi{\nabla
 _\nu}\psi-2{\nabla _\mu}{\nabla _\nu}\psi -{1\over
 2}g_{\mu\nu}(\nabla\psi )^2+2g_{\mu\nu}\Box\psi \right)\label{TPL}\ .\ee
We see that in the \PoL case the effective action looks much
simpler, being expressed in terms of only one auxiliary field.

Let us see which results do we get for the action \be \G _{PL}=\G
_g +\G _{1,PL}\ .\label{GPL}\ee For the auxiliary field $\psi$ we
have the same result (\ref{psires}), with the same value for the
integration constant $C=2\,{r_+^2-r_-^2\over r_+^2l}$. The regular
values of energy-momentum tensor are \be T_{uv} =2\k\
{(r^2-r_+^2)(r^2-r_-^2)(r^4+3r_+^2r_-^2)\over r^6l^4}\ ,\ee and
\be T_{uu} =T_{vv}=-2\k\
{(r^2-r_+^2)^2r_-^2(r^2(3r_+^2-r_-^2)-2r_+^2r_-^2)\over
r^6l^4r_+^2}\ .\ee The energy density is positive and has regular
behavior in the asymptotic region \be T_{tt}=4\k\
{(r^2-r_+^2)(r^6r_+^2-r^4r_-^2(4r_+^2-r_-^2)+r^2r_-^2r_+^2(6r_+^2+r_-^2)-5r_+^2r_-^2)
\over r^6l^4r_+^2}\ .\ee The asymptotic value of the energy
density in the locally flat frame is ${4\k\over l^2}$. \be
m(r)={2\over 3r^2r^6r_+^2}\Big(
2r_-(3r^4+3r^2(r_+^2+r_-^2)-5r_+^2r_-^2)+3r^3(r_+^2-r_-^2)^2\log{{r-r_-\over
r+r_-}}\Big)\ee \be\omega (r)={l(
-2r^2(3r_+^2+r_-^2)+8r_+^2r_-^2)\over r r_+^2 (r^2-r_-^2)} +{l(
3r_+^2+r_-^2)\over  r_+^2r_- }\log{{r+r_-\over r-r_-}}\ .\ee The
correction of curvature is $R_1=0$.

The results given above are particularly interesting because they
can be interpreted as corrections for $ \rm{AdS_2}$ black hole.
Namely, in the spinless case, the action (\ref{GPL}) describes
dilaton gravity with negative cosmological constant with the
quantum corrections produced by 2D minimally coupled scalar field.
The classical part of this action is Jackiw-Teitelboim model 2D
gravity \cite{jt}. The classical solution of the equations of
motion is $\rm{AdS}_2$ geometry
 \be
ds^2=-\,\left({r^2\over l^2}-lM \right)dt^2 + \left({r^2\over
l^2}-lM\right)^{-1}dr^2\ ,\label{ads2} \ee
  with the curvature $R_0=-{2\over l^2}$.
  $\rm{AdS}_2\times \rm{S}^2$ geometry appears as the near horizon
  geometry of the extermal \RN solution
   and it is analyzed in \cite{ss,fns00,fns01}.
  For the value of constant $C={ 2r_+\over l^2} $ we get the components of
  EMT:
  \be T_{uv}=2\k \ {r^2-r_+^2\over l^4}\ ,\ \ T_{uu}=T_{vv}=0\ ,\ \
  T_{tt}=4\k {r^2-r_+^2\over l^4}\ ,\ee
  for black hole in \HH state. $\rm{AdS}_2$ line element (\ref{ads2}) can be rewritten in the
null form: \be ds^2=-{lM\over \sinh^2\Big(\sqrt{{M\over
l}}{v-u\over 2}\Big)}dudv\ .\ee In this case $\rzv$ is
$$\rzv=-\sqrt{{l\over M}}{\rm Arc coth}{r\over \sqrt{l^3M}}\ .$$
 The Kruskal coordinates,
$$ U=-\sqrt{{l\over M}}e^{-\sqrt{M/l}\ u}\ ,V=\sqrt{{l\over
M}}e^{\sqrt{M/l}\ v}\ ,$$ are regular on the horizon,
$r=r_+=\sqrt{lM^3}$. The line element in these coordinates is
$$ds^2=-{4lM\over (1+MUV/l)^2}dUdV\ .$$ As we know, the \HH state
is the conformal state $\sket{UV} .$ It is easy to find the
components of EMT in this state using the law of transformation
the components of EMT from the Boulware, $\sket{uv}$ to \HH state,
$\sket{UV} .$ One can check that, performing this transformation
the previous result is obtained. Different vacuum states were, in
the framework of \RN geometry, discussed by Spradelin and
Strominger \cite{ss}. Fabbri, Navarro and Navarro-Salas
considered the one-loop corrections for evaporating $\rm{AdS}_2$
black hole \cite{fns00,fns01}, but again in the connection with
\RN geometry.

Now, we want to find the one-loop solution of this model. The
equations of motion take form: \be R=-{2\over l^2}\ , \nonumber\ee
  \be
g_{\alpha\beta} \Box \F -\nabla _\alpha \nabla _\beta \F-\F
g_{\a\b} {1\over l^2}=\frac{1}{2}T_{\a\b}\ ,\nonumber\ee where EMT
is given by (\ref{TPL}). This equations can be solved exactly. If
we assumed that the one-loop metric is given by (\ref{ads2}) than
we will obtain that the dilaton is given by \be
\F=\frac{r}{l}-2\k\ ,\ee for \HH state. It is interesting to note
that in the case of the Boulware vacuum (where $C=0$) there is
again exact solution: \be \F={r\over l}+ \k{r\over
r_+}\log{r+r_+\over r-r_+}\ .\ee The integration constants in
previous results are chosen in agreement with classical limit
$\k\to 0$. We see that the one-loop corrected metric is the AdS
black hole again - the quantum corrections neither change the
character of the space nor they produce the singularity at $r=0$.
\section{Conclusions}

In this paper we treated the one-loop corrections of dimensionally
 reduced BTZ black hole. We analyzed three types of effective actions,
 corresponding to different couplings of scalar matter (3D minimal,
 3D conformal and 2D minimal couplings).

 One of the main result is the analysis of the Unruh vacuum for reduced BTZ
 model. This state is defined demanding that EMT is regular on the
  future horizon.
   It has peculiar properties.

   The other point was to compare 2D
    reduced model with exact 3D results in the
   conformal case.
   Here we found that the corrections of geometry 2D reduced model
   is in a relatively good agreement
with Neumann boundary conditions for the scalar field. Let us note
that due to the ansatz (\ref{ans}) it is not possible to compare
the values of EMT directly.

Note that the energy density in the asymptotic region does not
obey Stefan-Boltzman law. This is not surprising if we keep in
mind that the Hawking radiation is not a free boson gas in this
region. The one-loop correction of entropy are logarithmic as it
is often the case.

Finally, in the last section we found exact result for JT model in
the case \HH and Boulware vacuum. In both cases there is not the
correction of ${\rm AdS_2}$ geometry in the quantum level. The
backreaction change dilaton field only. These two solutions will
be analyzed in the future publications.


\begin{thebibliography}{99}

\bibitem{btz} M. Ba\~ nados, C. Teitelboim and J. Zanelli, \PRL {\bf 69}, 1849
(1992).

\bibitem{bhtz} M. Ba\~ nados, M. Henneaux, C. Teitelboim and J. Zanelli, \PRD {\bf 48}, 1506
(1993).

\bibitem{ais} S. J. Avis, C. J. Isham and D. Storey, \PRD {\bf 18}, 3565
(1978).

\bibitem{ao} A. Achucarro and M. E. Ortiz, \PRD {\bf 48}, 3600
(1993).

\bibitem{mk} A. J. M. Medved and G. Kunstatter,
\PRD {\bf 63}, 104005 (2001).

\bibitem{fis} V.P. Frolov,
 W. Israel and S.N. Solodukhin, \PRD {\bf 54}, 2732 (1996).

\bibitem{bf} R. Balbinot and A. Fabbri, \PRD {\bf 59}, 044031
(1999).

\bibitem{brm} M. Buri\' c, V. Radovanovi\' c and A. Mikovi\' c,
\PRD {\bf 59}, 084002 (1999).

\bibitem{mv4} M. Buri\' c and V. Radovanovi\' c,
\PRD {\bf 63}, 044020 (2001).

\bibitem{fsz} V. Frolov, P. Sutton and A. Zelnikov, \PRD {\bf 61}, 024021
(2000).

\bibitem{bffnsz} R. Balbinot, A. Fabri, V. Frolov, P. Nicolini, P. Sutton
 and A. Zelnikov, \PRD {\bf 63}, 084028 (2001).

\bibitem{lo} L. Lifschytz and M. Ortiz, \PRD {\bf 49}, 1929
(1994).

\bibitem{sm} K. Shiraishi and T. Maki, \PRD {\bf 49}, 5286 (1994).

\bibitem{s} A. R. Steif, \PRD {\bf 49}, 585 (1994).

\bibitem{landau} L. D. Landau and E. M. Lifshitz, {\it Field
Theory}, Nauka, Moscow (1973).

\bibitem{mr4} A. Mikovi\' c and V. Radovanovi\' c,
Class. Quant. Grav.{\bf 15}, 827 (1998).

\bibitem{mr3} A. Mikovi\' c and V. Radovanovi\' c,
Nucl. Phys. B {\bf 504}, 511 (1997).

\bibitem{bh} R. Bousso and S.W. Hawking, \PRD{\bf 56}, 7788
(1997).

\bibitem{no} S. Nojiri and S.D. Odintsov,
Mod. Phys. Lett. A{\bf 12}, 2083 (1997).

\bibitem{klv} W. Kummer, H. Leibl and D. V. Vassilevich, Mod. Phys.
Lett. A{\bf12}, 2683 (1997).

\bibitem{kv99} W. Kummer and D. V. Vassilevich, Analen.Phys. {\bf 8}, 801
(1999).

\bibitem{klv98} W. Kummer, H. Leibl and D. V. Vassilevich, \PRD {\bf 58}, 108501
(1998).

\bibitem{klv99} W. Kummer, H. Leibl and D. V. Vassilevich, \PRD {\bf 60}, 084021
(1999).

\bibitem{lmr} F. Lombardo, F. D. Mazitelli and J. G. Russo, \PRD {\bf 59}, 064007
 (1999).

\bibitem{no99} S. Nojiri and S. D. Odinstov, Int. Journal of Mod. Phys.
A{\bf 15}, 989 (2000).

\bibitem{nooo99} S. Nojiri, O. Obregon, S. D. Odinstov and K. E. Osertin,
 \PRD {\bf 60}, 024008 (1999).

\bibitem{cf} S. M. Christensen and S. A. Fulling, \PRD {\bf 15},
2088 (1977).

\bibitem{w} R. M. Wald, \PRD {\bf 48}, 3427 (1993).

\bibitem{mv3} M. Buri\' c and V. Radovanovi\' c, Class. Quant. Grav. {\bf
17}, 33 (2000).

\bibitem{ss} M. Spradlin and A. Strominger, JHEP {\bf 9911}, 021
(1999).

\bibitem{fns00} A. Fabbri, D. J. Navarro and J. Navaro-Salas,
Phys. Rev. Lett. {\bf 85} (2000) 2434.

\bibitem{fns01} A. Fabbri, D. J. Navarro and J. Navaro-Salas,
Nucl. Phys. B {\bf 595}, 381 (2001).

\bibitem{mz} C. Martinez and J. Zanelli, \PRD {\bf 55}, 3642
(1997).

\bibitem{jt} R. Jaciw, in "Quantum Theory of Gravity", edited by
S.M. Christensen (Hilger, Bristol., 1984), p. 403; C. Teitelboim,
in op. cit., p327.


\end{thebibliography}
\end{document}